\documentclass[prl,aps,twocolumn,amsfonts,showpacs,superscriptaddress,nofootinbib]{revtex4-1}

\pdfoutput=1

\usepackage{epsfig} 
\usepackage{psfrag}
\usepackage{amsmath}
\usepackage{amssymb}
\usepackage{color}
\usepackage{graphicx}
\usepackage{subfigure}
\usepackage{hyperref} 
\usepackage{enumitem}

\newcommand{\be}{\begin{equation}}
\newcommand{\ee}{\end{equation}}
\newcommand{\bea}{\begin{eqnarray}}
\newcommand{\eea}{\end{eqnarray}}

\begin{document}

\title{Universal spatial structure of nonequilibrium steady states}

\author{Julian Sonner} 
\affiliation{Department of Theoretical Physics, University of Geneva, 24 quai Ernest-Ansermet, 1214 Gen\`eve 4, Switzerland}  

\author{Benjamin Withers} 
\affiliation{Department of Theoretical Physics, University of Geneva, 24 quai Ernest-Ansermet, 1214 Gen\`eve 4, Switzerland}  

\begin{abstract}
  We describe a large family of nonequilibrium steady states (NESS) corresponding to forced flows over obstacles. The spatial structure at large distances from the obstacle is shown to be  universal, and can be quantitatively characterised in terms of certain collective modes of the strongly coupled many body system, which we define in this work. In holography, these modes are spatial analogues of quasinormal modes, which are known to be responsible for universal aspects of relaxation of time dependent systems. These modes can be both hydrodynamical or non-hydrodynamical in origin.
The decay lengths of the hydrodynamic modes are set by $\eta/s$, the shear viscosity over entropy density ratio, suggesting a new route to experimentally measuring this ratio.
We also point out a new class of nonequilibrium phase transitions, across which the spatial structure of the NESS undergoes a dramatic change, characterised by the properties of the spectrum of these spatial collective modes. 
\end{abstract}

\date{May 2017}

\pacs{}

\maketitle
Equilibrium many-body systems are known to exhibit universal behaviour, as famously exemplified by their critical phenomena near second-order phase transitions. These are characterised by a small number of universal modes that scale according to computable critical exponents and leave their imprint on macroscopic physical properties of the system.

This state of affairs contrasts with the situation when such systems are not in equilibrium \cite{hohenberg1977theory}, in which case universal results are few and far between. Determining the physical characteristics of such a system is typically strongly situation dependent. A notable and, for our purposes, illustrative exception is the dynamical crossing of a second-order phase transition at a finite rate $\tau_{\rm Q}$. In this case, as proposed by Kibble \cite{Kibble:1976sj} and Zurek \cite{Zurek:1985qw}, the number of topological defects that form in the broken symmetry phase is given in terms of a scaling law, whose input is a small set of universal modes characterising the theory. The exact details of the quench through the transition are unimportant, only the rate of approach to the critical point enters into the scaling law  \cite{Zurek:1985qw}. 

Given the success of the KZ mechanism \cite{delCampo:2013nla}, and the recent experimental interest it has created, for example \cite{navon2015critical,chomaz2015emergence}, one may ask whether other  scenarios exist that are able to strongly constrain out of equilibrium dynamics using a small set of universal collective modes, leaving an imprint on the macroscopic spatial structure of the system.

\begin{figure}
\includegraphics[width=\columnwidth,clip]{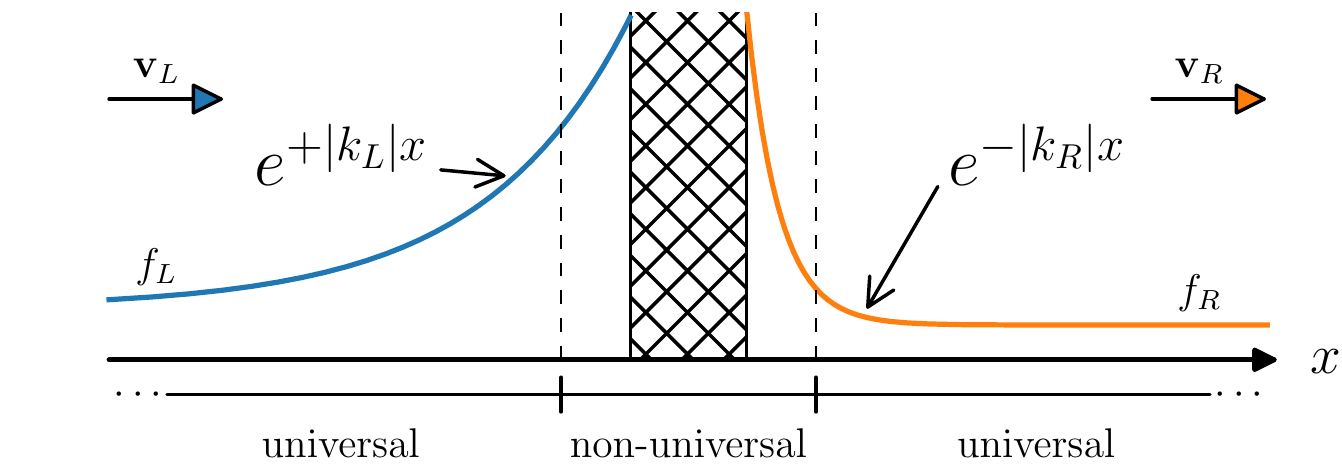}
\caption{Schematic representation of the NESS considered, showing the imprint of spatial collective modes which describe the return to equilibrium far from an obstacle. }
\label{Fig.Schema}
\end{figure}

In this work we consider a large class of nonequilibrium steady states (NESS), which are set up as follows: consider a (quantum critical) many body system forced to flow over an obstacle. This gives rise to a strong non-linear disturbance in the vicinity of the obstacle, while the flow far from it on either side is simple with a constant velocity ${\bf v}_{\rm L}$ on the left and  ${\bf v}_{\rm R}$ on the right (see Fig. \ref{Fig.Schema}). 
One then wants to know what the steady state looks like at large distances, in other words how the strongly non-linear behaviour around the obstacle relaxes spatially toward its asymptotic values. This is a difficult problem, in general out of technical reach of current methods. The AdS/CFT correspondence gives rise to a powerful computational framework particularly in the nonequilibrium setting. Indeed this approach has been used to elucidate the temporal equilibration\footnote{Furthermore, previous studies of holographic NESS include current driven \cite{Karch:2010kt,Sonner:2012if,Nakamura:2013yqa,Kundu:2013eba} as well as heat-driven \cite{Chang:2013gba,Bhaseen:2013ypa} cases.} of strongly coupled plasmas \cite{Chesler:2008hg,Heller:2012km} and superfluids \cite{Bhaseen:2012gg}. In each case, the late-time behaviour is very accurately predicted by the spectrum of low-lying quasinormal modes (QNM) \cite{Kovtun:2005ev}, whose relevance to thermalization was first pointed out in \cite{Horowitz:1999jd}.

In this paper we use holography to explicitly find the full non-linear solution for certain strongly coupled theories, where the dual solutions are given by black holes without Killing horizons. The spatial structure in all examples is indeed universal, and characterised by a stationary version of QNMs\footnote{Modes of this kind have been studied in holography in a variety of other contexts \cite{Csaki:1998qr, Amado:2007pv, Maeda:2009wv, Khlebnikov:2010yt, Sonner:2014tca}.}, which we define and obtain in a few illustrative examples. For a given choice of asymptotic flow velocity, $v = {v}_{\rm L}$ or ${v}_{\rm R}$, these modes form a discrete set of purely imaginary wavenumbers $k(v)$ and the leading mode, i.e. the one with the smallest $|{\rm Im}k|$ can be hydrodynamical or non-hydrodynamical, and  will be denoted $k_*$. The relaxation towards the asymptotic flow happens at the exponential rate $\propto e^{-{\rm Im}k_* x}$, so that the relaxation towards the right boundary value corresponds to a mode with ${\rm Im}k_* > 0$, while the left mode has ${\rm Im}k_* < 0$. A drastic reorganisation of the spatial structure of the NESS occurs whenever a dominant mode crosses the real axis for a certain critical velocity $v_{\rm c}$. In this case, as $v\rightarrow v_{\rm c}^-$ the downstream spatial relaxation rate will tend toward zero, only to be, for $v>v_{\rm c}$, dominated by the previously subleading mode. The upstream spatial relaxation rate undergoes a similar transition as $v$ is decreased through $v_{\rm c}$. This reorganisation of the spatial structure constitutes a new nonequilibrium phase transition, and we conjecture that transitions of this form exist in systems outside of holography. Indeed we provide examples of such transitions purely from the point of view of hydrodynamics. 

The physical setup considered in this work should be regarded as a spacelike version of a quench~\cite{Figueras:2012rb}. Instead of switching on a source at some time $t_0$ and then asking about the temporal relaxation towards a new equilibrium, we consider an obstacle (modelled by a source) at some spatial location $x_0$ and asking about the spatial relaxation towards the asymptotic equilibrium. In both cases the asymptotic physics is fully universal and determined by a spectrum of discrete collective modes of the system. The importance of QNMs in holography cannot be overstated, and attempts are being made to define and explore them beyond AdS/CFT \cite{Heller:2015dha}. Here we point out that an equally rich and universal story is present when considering NESS, opening the particularly exciting possibility to access these modes via measurements of the spatial structure of driven critical systems in the lab. In particular, for modes which are hydrodynamic in origin the spatial decay rate (in units of the temperature) depends directly on the shear viscosity in units of the entropy density, $\eta/s$. This applies for any system with an effective hydrodynamic description, greatly extending the scope beyond holography and raising the interesting possibility of an experimental measurement of $\eta/s$ using the spatial structure of NESS. To this end we note that recent experiments have demonstrated the presence of hydrodynamic electron flow in PdCoO$_2$ \cite{moll2016evidence}, as well as graphene \cite{crossno2016observation}.

\textbf{Relativistic hydrodynamics in $d$ dimensions.}
Hydrodynamics describes a wide class of systems in the form of a universal theory which arises in a long wavelength limit.
In this section we construct the spatial collective modes that appear in this effective theory.
We stress that whilst hydrodynamics does contain certain spatial collective modes, there can be additional `higher' modes in a more complete theory that do not exist in the hydrodynamic limit. This is the case for holography, discussed in the next section.

To first order the Landau frame stress tensor is
\be
T^{\mu\nu} = \varepsilon u^\mu u^\nu + p \Delta^{\mu\nu} - \eta \sigma^{\mu\nu} - \zeta \Delta^{\mu\nu} \partial\cdot u + O(\partial)^2
\ee
subject to the conservation equations, $\partial_\mu T^{\mu\nu}=0$. $u^\mu$ is a timelike unit-normalised $d$-velocity field, while $\Delta^{\mu\nu} = \eta^{\mu\nu} + u^\mu u^\nu$ projects orthogonal to $u^\mu$. $\eta$ and $\zeta$ are the shear and bulk viscosities. The shear tensor is given by $\sigma^{\mu\nu} \equiv 2 \Delta^{\mu\rho}\Delta^{\nu\sigma} \left(\partial_{(\rho}u_{\sigma)} - \tfrac{1}{d-1}\eta_{\rho\sigma} \partial\cdot u\right)$. 

To find the collective modes, we solve the conservation equations for linear perturbations about a long-range equilibrium state characterised by $\varepsilon, p$ and a $(d-1)$-velocity, ${\bf v}$, such that $u^\mu = \gamma (1,{\bf v})$ where $\gamma = 1/\sqrt{1-{\bf v}\cdot{\bf v}}$. The perturbations we seek are of the form, $\varepsilon(x^\mu) = \varepsilon + \delta\varepsilon\, e^{i k_\sigma x^\sigma}$ with similar expressions for $p(x^\mu)$ and $u^\mu(x^\nu)$, all of which are time independent in the laboratory frame, i.e. $k_\mu = (0, {\bf k})$. Energy conservation immediately gives $\delta \varepsilon = - (\varepsilon+p) k\cdot \delta u /(k\cdot u)$, and for a speed of sound $c_s$ we also write $\delta p = c_s^2 \delta \varepsilon$. Thus, the remaining unsolved conservation equations determine $\delta u^\mu$, which are either transverse or longitudinal with respect to the obstacle. 
Transverse perturbations, $k\cdot \delta u_T = 0$ (and hence $\delta \varepsilon_T = \delta p_T = 0$), obey the dispersion relation
\be
k = -i \frac{\varepsilon + p}{\eta} v \cos\theta + O(k^2), \label{hydroshear}
\ee
where we denote $v=|{\bf v}|, k = \sqrt{{\bf k}\cdot{\bf k}}$, and ${\bf v}\cdot{\bf k} = vk\cos\theta$, obtained by solving for $v$ order-by-order in small $k$, and then inverting.
Despite being time independent, this mode is related to the usual shear diffusion pole. Specifically, by if we perform a Lorentz transformation to the rest frame of the fluid where the wavevector picks up a frequency $k^\mu = (\omega, {\bf q})$, at this order these quantities obey a dispersion relation of the form $\omega = -i D q^2$ with diffusion constant $D = \frac{\eta}{\varepsilon+p}$. Note however that $q$ here is imaginary.
Next the longitudinal sector,  $\delta u^\mu = \delta u_L \Delta^{\mu\nu}k_\nu$, has a dispersion relation,
\be
k = -i \frac{\varepsilon+p}{\frac{d-2}{d-1}\eta + \frac{1}{2}\zeta} \frac{\sqrt{1-v_0^2}\cos\theta}{(1-(v_0\sin\theta)^2)^2}(v\mp v_0) + O(k)^2,\label{hydrosound}
\ee
where $v_0 \equiv  c_s \sec\theta/\sqrt{1+ (c_s\tan\theta)^2}$.
Similarly this mode is related to sound; in the rest frame of the fluid it obeys the dispersion relation $\omega = \mp c_s q - \frac{i}{2} \frac{\frac{d-2}{d-1}2\eta +  \zeta}{\varepsilon + p} q^2$, but again note $q$ is imaginary.

The appearance of $\eta, \zeta$ in $k(v)$ suggests a new route to their measurement (as well as other transport coefficients which appear at higher orders in $k$)-- by measuring the long range spatial structure of NESS  in the laboratory. Specifically, using $\varepsilon + p = Ts$ we see that $k/T$ in \eqref{hydroshear} depends only on $\eta/s$ and parameters of the flow ($v,\theta$), whilst \eqref{hydrosound} depends additionally on $\zeta/s$ and $c_s$. The preceding analysis relies only on universal properties of hydrodynamics, and is thus independent of holographic duality, to which we turn next.

\textbf{Holography for CFT$_3$.}
Moving to a complete theory allows us to construct a complete spectrum -- hydrodynamic and otherwise -- as well as demonstrate its role in explicitly constructed NESS. 
Holography is a tool which makes such computations for a CFT$_d$ accessible through a soluble gravity dual in $d+1$ spacetime dimensions.

As before we construct the spectrum of spatial collective modes by linearly perturbing the equilibrium solution reached far from the obstacle. In this case the equilibrium configuration is given by a dual bulk gravity solution, the Schwarzschild black brane metric, which is boosted along a planar horizon direction by an amount corresponding to the asymptotic flow 3-velocity, $u^\mu$.\footnote{Note that one could also consider the approach to different equilibrium states, for instance those with charge, superconductors, insulators, etc. Such states would correspond to more exotic black brane solutions, some with extra matter fields in the bulk.} We adopt ingoing Eddington-Finkelstein coordinates,
\be
ds_{Schw.}^2 = \frac{1}{z^2}\left(-f(z) (u_\mu dx^\mu)^2 + 2 u_\mu dx^\mu dz + \Delta_{\mu\nu}dx^\mu dx^\nu\right). \label{metricS}
\ee
The conformal boundary is located at $z=0$ and the metric function $f(z) = 1-z^3/z_h^3$ vanishes at $z=z_h$, the black hole event horizon.

The spatial collective modes are linear perturbations of this metric. We work with the gauge-fixed ansatz for perturbations,
\be
\delta g_{ab}(z,x^\mu) dx^a dx^b = z^{-2}h_{\mu\nu}(z) e^{i  k_\sigma x^\sigma} dx^\mu dx^\nu,
\ee
which give rise to a set of coupled ODEs in $z$ for the Einstein equations. The $h_{\mu\nu}$ naturally organise into longitudinal and transverse channels with respect to the obstacle. We first define $n^\mu \propto \epsilon^{\mu\nu\rho}k_\nu u_\rho$ so that $n\cdot u = n\cdot k = 0$. Schematically we have coupled ODEs for the perturbations $h_{uu}, h_{uk}, h_{kk}, h_{nn}$ in the longitudinal channel and for $h_{un}, h_{kn}$ in the transverse channel.

To complete the holographic prescription for the modes we must specify boundary conditions. At $z=0$ we require that no external sources are turned on. 
This computation is reminiscent of a QNM calculation where the boundary condition at the event horizon is ingoing, equivalent to regularity on the future event horizon. Here, in the laboratory frame, the spatial collective modes are time independent by construction, so an ingoing condition cannot apply. We define the modes to be those which are regular on the future event horizon.\footnote{In particular, for the black holes dual to a NESS, we may have access only to the future event horizon. Note that for the perturbations of Schwarzschild considered here, the boundary condition coincides with an ingoing one in other inertial reference frames, such as the rest frame of the flow.}
In addition we must select the mode which is regular at infinity, so that on the right hand side of the obstacle we require ${\rm Im}k\geq 0$, and on the left, ${\rm Im}k\leq 0$. Of course, a right hand side mode in isolation is not regular because it blows up as $x\to -\infty$, but such modes can appear on the right hand side of a regular NESS. 

\begin{figure}
\vspace*{0.2cm}
\includegraphics[width=\columnwidth]{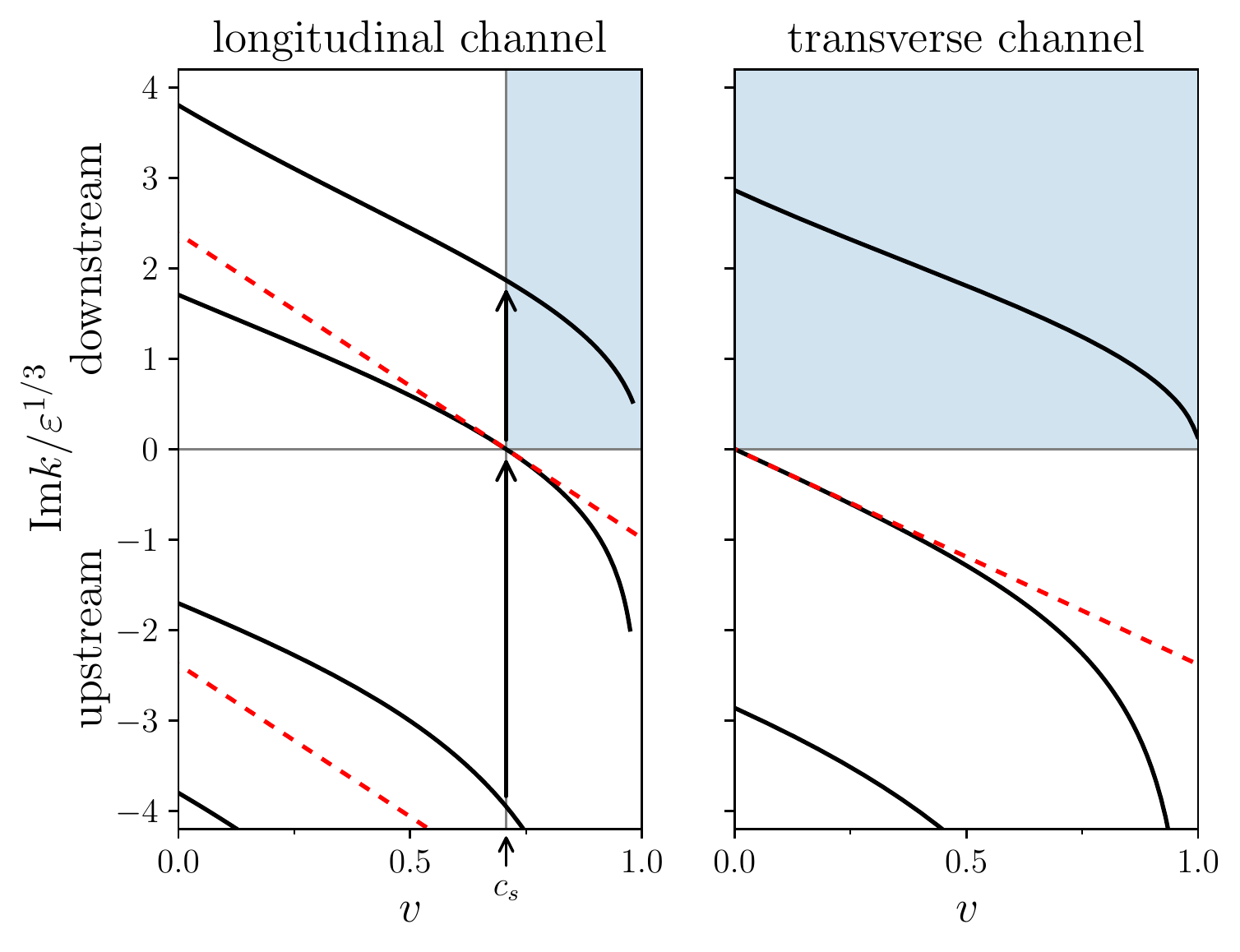}
\vspace*{-0.4cm}
\caption{The discrete spectrum of spatial collective modes as a function of asymptotic flow velocity, $k(v)$, for a CFT$_3$, computed holographically using stationary perturbations of boosted Schwarzschild-AdS$_4$. Here we show the case of flow incident angle $\theta = 0$ (black). There is a $(v,k)\to(-v,-k)$ symmetry which connects some of the modes shown through $v=0$.  Also shown is the conformal relativistic hydrodynamic spectrum (red dashed) valid to first order in small $k$. All modes found have ${\rm Re}k=0$. On the downstream side, for some flow velocities $v$ there are no modes of hydrodynamic origin (blue shaded region). In the longitudinal channel there is a phase transition as the velocity is increased through $c_s$ (arrows) giving rise to discontinuities in $k_*$.}
\label{Fig.Spectrum}
\end{figure}

We solve these equations numerically by shooting, subject to the boundary conditions outlined above. The leading (i.e. longest range) parts of the resulting spectrum are displayed in Fig. \ref{Fig.Spectrum}. For comparison we also show the modes obtained in the first-order hydrodynamic approximation, with appropriate transport coefficients $\eta = s/(4\pi), \zeta = 0$ and $c_s = 1/\sqrt{2}$. All modes found have ${\rm Re}k=0$. As previously advertised the holographic theory contains additional modes that are not present in hydrodynamics and, crucially, for some $v$ these non-hydrodynamic modes give the dominant long distance contribution. 

A new nonequilibrium phase transition is also visible in Fig. \ref{Fig.Spectrum}. In the longitudinal channel, as $v$ is increased through $c_s$, there is a sudden change in the dominant mode, $k_*$, on either the upstream or downstream side. For instance, on the downstream side the hydrodynamic mode decay length becomes ever longer as $v$ is increased, and becomes suddenly dominated by a short non-hydrodynamic mode once $v>c_s$. On the upstream side the transition takes place between modes that are hydrodynamic in origin.

\textbf{Holography for CFT$_2$ and CFT$_\infty$.}
In low and high spacetime dimension analytic treatment of the spatial collective modes becomes possible.
For $d=2$ equilibrium is given by the BTZ black brane. For a scalar field perturbation about the zero velocity background there is a discrete set of modes labelled by $n\in \mathbb{Z}$, whose dispersion relations are given by $\omega = \pm q - i 4\pi T(\frac{\Delta}{2}+n)$ \cite{Cardoso:2001hn,PhysRevD.64.064024} where $T$ is the Hawking temperature of the black hole and $\Delta$ is the dimension of the operator dual to the scalar. Exploiting Lorentz invariance to reach the modes of interest, i.e. time independent modes for a background with velocity $v$, we pick $\omega =- \gamma k v$, $q = \gamma k$, obtaining
\be
k = i \frac{4\pi T}{\gamma(v\pm 1)} \left(\frac{\Delta}{2} + n\right),
\ee
where ${\rm Re}k=0$ and -- comparing to \eqref{hydroshear}, \eqref{hydrosound} -- a suggestive factor of $4\pi T$, despite $\eta$ not being defined in $d=2$.
In the limit $d\to\infty$ there is a decoupled sector of perturbations which are supported in a near horizon region, corresponding to modes with $\omega, q \sim d^0$ \cite{Asnin:2007rw,Emparan:2013moa,Emparan:2013xia}. These can be constructed analytically \cite{Emparan:2015rva}. Once more using Lorentz invariance an appropriate choice of $\omega, q$ gives ${\rm Re}k=0$. For small $k$ these modes match the large $d$ limit of the hydrodynamic modes computed earlier.

\textbf{Nonlinear holographic NESS construction.}
In the previous sections we constructed individual spatial collective modes. 
Here we show that these modes govern the behaviour the NESS far from the obstacle by explicitly constructing a NESS and checking the asymptotics.
These are given holographically by families of black branes with non-Killing horizons, in which the obstacle is provided by $x$-dependent deformations of the CFT metric, $\gamma_{\mu\nu}$, i.e.
\be
\gamma_{\mu\nu} = \eta_{\mu\nu} + s_{\mu\nu}(x). \label{boundarymetric}
\ee
We consider sources whose components are Gaussian centred, on $x=0$. 
The details of the obstacle are not important, as the spectrum of collective modes is a property of the theory itself. We only have to ensure that the obstacle excites the part of the spectrum we are interested in. The source terms in \eqref{boundarymetric} can act as a source for shear, and we allow for velocity components transverse to the obstacle. A NESS corresponding to a CFT flowing over a step function obstacle was studied in ideal hydrodynamics by \cite{Fischetti:2016tek}.

Our construction proceeds numerically based on the method of \cite{Figueras:2012rb} which formulates the stationary gravitational problem such that the bulk coordinates penetrate the future event horizon. 
As emphasised in \cite{Figueras:2012rb}, one must supply enough data in the form of boundary conditions to fix all the moduli of the corresponding flow. In addition to $\varepsilon, v$ of \cite{Figueras:2012rb}, we fix a third modulus, $\theta$, the asymptotic incident angle of the flow. In general there is refraction and $\theta_L \neq \theta_R$.

We have constructed solutions which are asymptotically subsonic-to-subsonic, as well as supersonic-to-supersonic, with and without transverse flow. For these solutions we seek local fluid variables by using the field theory stress tensor, $\left<T_{\mu\nu}\right>$ obtained using holographic renormalisation \cite{deHaro:2000vlm}. 
We solve the following eigenvalue problem at each point on the boundary,
\be
\left<T_{\mu\nu}\right> U^\mu = -\varepsilon U_\nu, \qquad \gamma_{\mu\nu}U^\mu U^\nu = -1 \label{evalproblem}
\ee
for the three undetermined pieces of $\varepsilon, U^\mu$. Asymptotically on the left or right these are the moduli of the solution, i.e. asymptotically $U^\mu = \gamma(v)(1,v \cos\theta, v \sin\theta)$. 

\begin{figure}
\includegraphics[width=\columnwidth,clip]{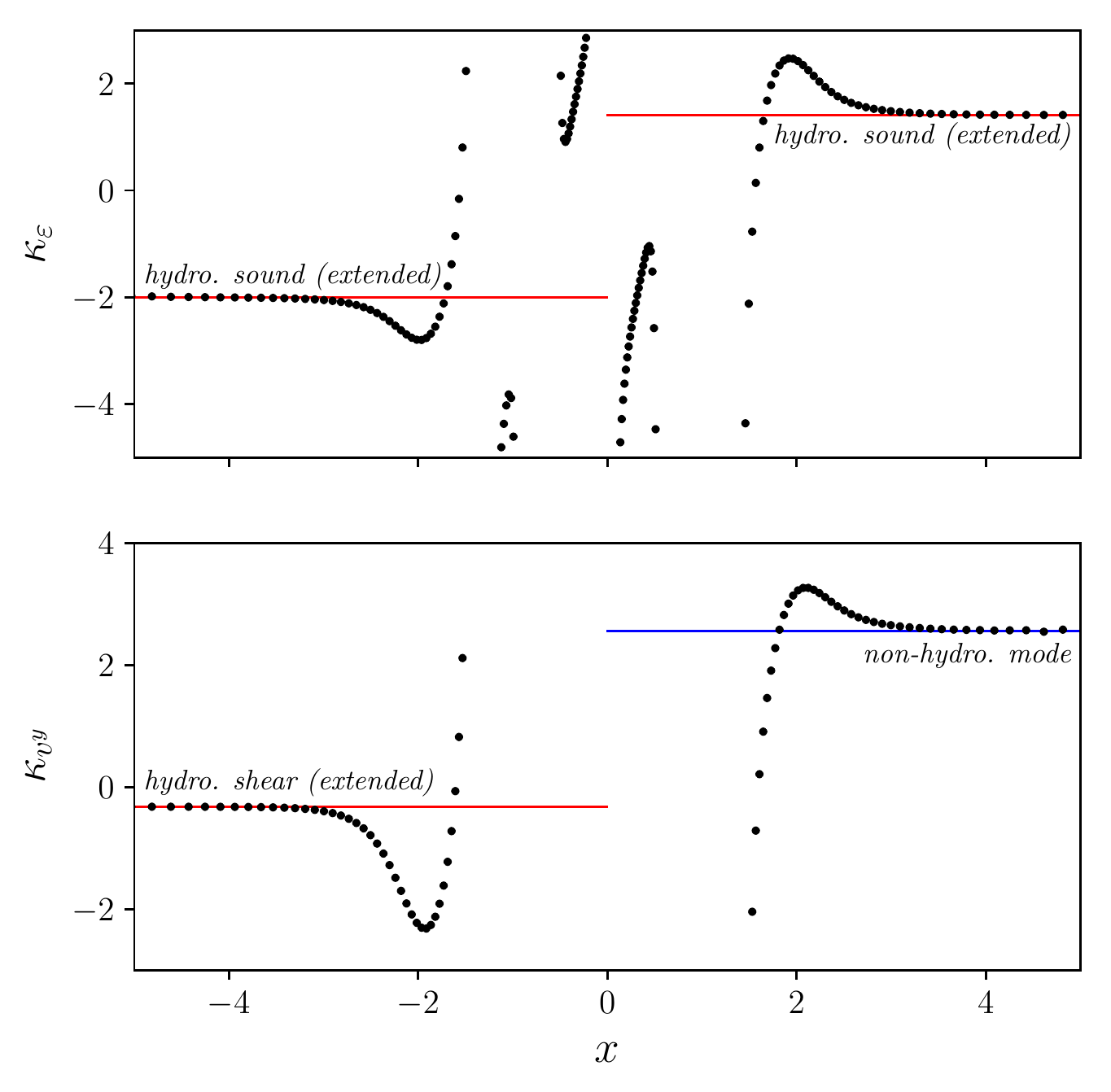}
\caption{Asymptotically subsonic-to-subsonic NESS, with finite transverse velocity. We show (with black circles) $\kappa_\varepsilon$ for the longitudinal channel (\emph{upper panel}) and $\kappa_{v^y}$ for the transverse channel (\emph{lower panel}), as defined in \eqref{kappadef}. Also shown are the values of ${\rm Im}k/\varepsilon^{1/3}$ for the spatial collective modes, computed directly given the left or right moduli of the asymptotic equilibrium. The red solid lines are continuously connected to the hydrodynamic modes labelled, whilst the blue solid line is a non-hydrodynamic mode.}
\label{Fig.ShearFit}
\end{figure}

To check for the presence of the collective modes we note some quantity $f$ in the channel of interest will take the form $f = C + A_k e^{-{\rm Im}k\,x}$. To numerically extract the value of $k$ we then compute
\be
\kappa_f(x) = -\frac{1}{\varepsilon^{1/3}}\frac{\partial_x^2 f}{\partial_x f} \label{kappadef}
\ee
and then ${\rm Im}k/\varepsilon^{1/3} = \lim_{x\to\pm\infty} \kappa_f(x)$. To illustrate we use an example where a mode of non-hydrodynamic origin is dominant. One place this occurs is in the transverse channel, downstream in a subsonic flow (as we may predict from the spectrum of Fig. \ref{Fig.Spectrum}). We give an example of this flow in Fig. \ref{Fig.ShearFit} where we show $\kappa_\varepsilon$ and $\kappa_{v^y}$ where $v^y  = U^y/U^t$. These quantities display excellent agreement with the longest range spatial collective mode obtained by direct construction, confirming the expectation that the spatial collective modes determine the long distance behaviour of the nonlinear NESS.\footnote{Note that $\kappa$ becomes precision limited for larger values of $|x|$, as expected due to the exponential decay of the mode amplitudes with $|x|$.}

\begin{figure}
\includegraphics[width=\columnwidth,clip]{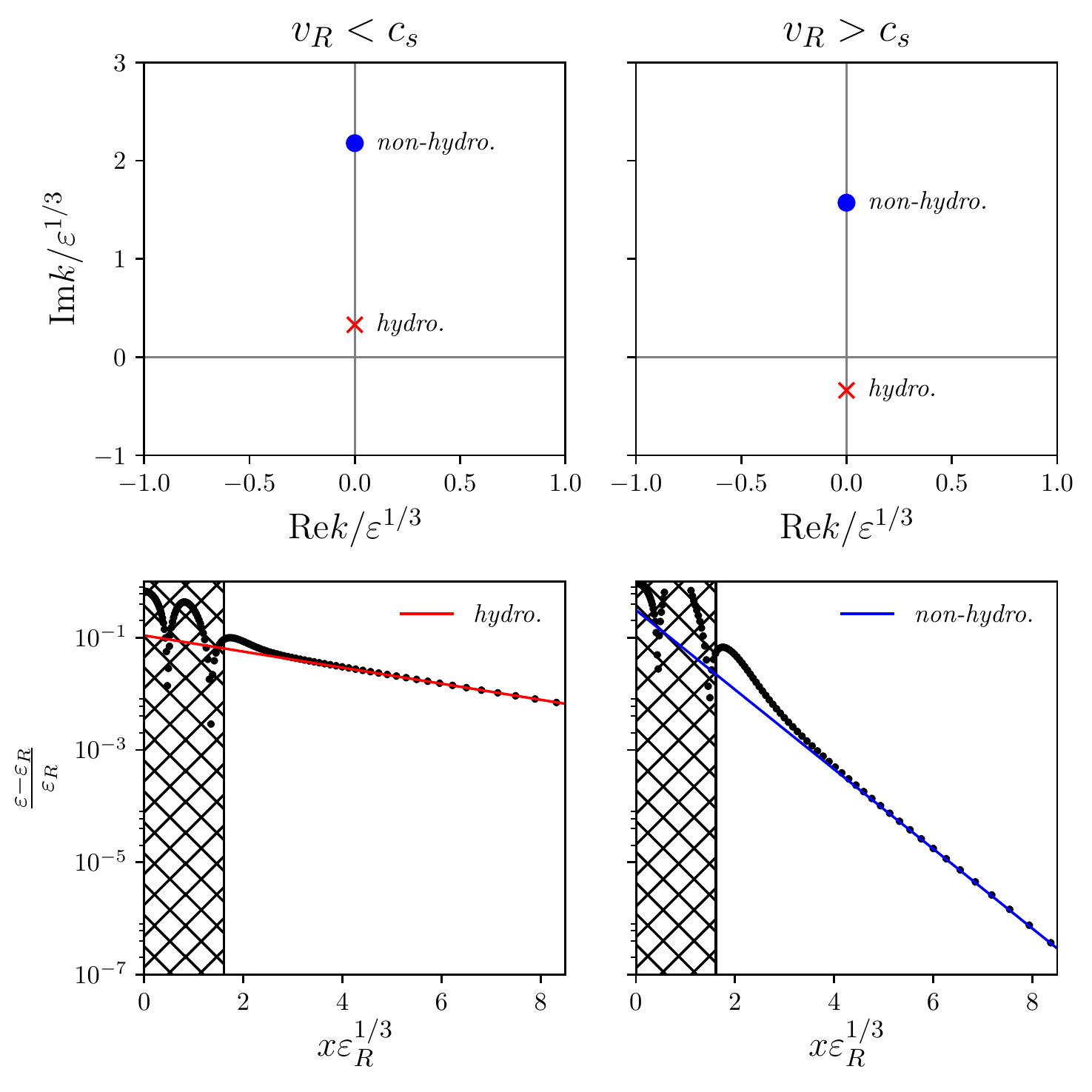}
\caption{Demonstration of the new nonequilibrium phase transition on the downstream, right hand side of a NESS, from $v_R<c_s$ (\emph{left column}) to $v_R>c_s$ (\emph{right column}). 
\emph{Top row:} locations of the spatial collective modes at these $v_R$ in the complex $k$ plane, displaying one mode of hydrodynamic origin (red x) and one non-hydrodynamic mode (blue circle). 
\emph{Bottom row:} Spatial profile of $\varepsilon$ on the right hand side of a NESS (black circles) together with an amplitude-fit collective mode from the spectrum above with the longest decay length (solid lines).}
\label{Fig.Transition}
\end{figure}

Finally, we turn to a demonstration of the proposed nonequilibrium phase transitions in the longitudinal channel at $v=c_s$.  In Fig. \ref{Fig.Transition} we consider the downstream, right hand side of a NESS in two cases, $v_R < c_s$ and $v_R>c_s$. In each case we show the spatial decay of $\varepsilon$ and the longitudinal collective mode spectrum on the complex-$k$ plane. Beginning with $v_R<c_s$, the long range behaviour is governed by the smaller ${\rm Im}k>0$ mode, as the plot of $\varepsilon$ indicates. As $v_R$ is increased, this mode descends down the imaginary-$k$ axis and crosses the real axis at $v_R=c_s$. For $v_R>c_s$ this mode is in the lower half plane, no longer decays as $x\to+\infty$, and so it can no longer appear on the right hand side of a regular NESS. The behaviour of $\varepsilon$ is thus suddenly dominated by the second, non-hydrodynamic mode which is now the longest range contribution.

\textbf{Discussion.}
We have defined and constructed `spatial collective modes' which, as we have argued, describe the universal spatial relaxation to equilibrium at large distances in a wide class of NESS. 
In the hydrodynamic limit the decay length of the modes depend directly on $\eta/s$, suggesting a new route to its experimental measurement.
 The often delicate issue of heating in NESS (see e.g. \cite{green2005nonlinear,green2006current}) here is elegantly sidestepped, since the spatial pattern of the heat flow itself is universal and predicted by our mechanism. We have constructed explicit examples of non-Killing black holes in holography which confirm the role played by these modes, and demonstrated novel nonequilibrium phase transitions resulting from a reorganisation of their spectrum.   It is our hope that these modes, which may be viewed as the spatial analogues of QNMs, provide fruitful targets for further theoretical and experimental work on nonequilibrium steady states.

\begin{acknowledgments}
We thank A. del Campo, J. Gauntlett, A. Green, I. Novak, G. Policastro, K. Schalm, U. Schollw\"ock and T. Wiseman for discussions.
This research is supported by the Fonds National Suisse de la Recherche Scientifique (FNS) under grant number 200021 162796 and by the NCCR 51NF40-141869 ``The Mathematics of Physics" (SwissMAP).
\end{acknowledgments}

\bibliographystyle{utphys}
\bibliography{NESS}{}

\section{Supplemental material}
\subsection{Spatial collective modes of Schwarzschild}
The metric perturbations considered are of the form,
\be
\delta g_{ab}(z,x^\mu) dx^a dx^b = z^{-2}h_{\mu\nu}(z) e^{i  k_\sigma x^\sigma} dx^\mu dx^\nu.
\ee
on a background with velocity 3-vector $u^\mu$. 
We first form an orthogonal basis, $(u, k_\perp, n)$, where $k_\perp^\mu = \Delta^{\mu}_{~\nu}k^\nu$ and where $n^\mu = N\epsilon^{\mu\nu\rho}k_\nu u_\rho$ with some normalisation $N$. We then write the metric perturbations in this basis, i.e.
\bea
h_{\mu\nu} &=& h_{un} u_{(\mu} n_{\nu)} + h_{kn} k^\perp_{(\mu} n_{\nu)}\\
&&+ h_{uu} u_{(\mu} u_{\nu)} + h_{kk} k^\perp_{(\mu} k^\perp_{\nu)}+ h_{uk} u_{(\mu} k^\perp_{\nu)} + h_{nn} n_{(\mu} n_{\nu)}\nonumber
\eea
where the coefficients are functions of $z$. The equations of motion then naturally separate by odd or even parity under $n\to -n$. 

It is convenient to consider an explicit frame of reference. In the laboratory frame we have purely spatial $k^\mu = (0, {\bf k})$ with $u^\mu = \frac{1}{\sqrt{1-v^2}}(1,{\bf v})$. In the fluid rest frame we have $k^\mu = (\omega, {\bf q}) =  (-\gamma {\bf v}\cdot {\bf k}, {\bf k}+ \frac{\gamma-1}{v^2} ({\bf v}\cdot {\bf k}) {\bf v})$ with $q=|{\bf q}|$. We find the equations to be more manageable when expressed in fluid rest frame variables $\omega, {\bf q}$. In the odd (transverse) sector we have equations of motion,
\bea
q^2(f\partial_z + i \omega)h_{kn} &=& (\omega\partial_z + i q^2) h_{un}, \nonumber\\
(z f \partial_z^2 + (f+2i\omega z -3)\partial_z - 2 i\omega) h_{kn} &=& i (z \partial_z - 2) h_{un}. \nonumber
\eea
Near the horizon at $z=z_h$ the solution takes the following general form, 
\bea
h_{un} &=& a(z-z_h)^0(1 + \ldots) + c(z-z_h)^{\frac{2i\omega z_h}{3}}(0 + \ldots)\nonumber\\
h_{kn} &=& b(z-z_h)^0(1 + \ldots) + c(z-z_h)^{\frac{2i\omega z_h}{3}}(1+ \ldots)\nonumber
\eea
with three undetermined pieces of data $a, b, c$.
The ellipses denote terms of higher order in $(z-z_h)$ which are completely determined once the data here are specified.
For generic ${\bf v}\cdot {\bf k}$ the $c$-terms are not regular at the horizon, and so we set $c=0$. This coincides also with an ingoing boundary condition in the fluid rest frame.
Near the boundary, once sources are turned off, the fields take the following form
\bea
h_{un} &=& A q^2 z^{3} + \ldots\nonumber\\
h_{kn} &=& A \omega z^{3} + \ldots\nonumber
\eea
with a single undetermined piece of data $A$. The ellipses denote additional powers of $z$ with coefficients that contain no new data. 
Without loss of generality we may fix $z_h =1$, and we may set one of $a,b,A$ by linearity, say, $b$.
We have a system of equations with total differential order $3$ and for fixed $v$ and $\theta$ we have the remaining three parameters $(a,A,k)$. Thus we expect to find discrete solutions at fixed $v$ and $\theta$. These solutions are constructed numerically using a standard shooting method.

A similar analysis applies to the (lengthier) equations in the longitudinal sector, which we omit here.

\subsection{Numerical construction of non-Killing black branes}
Black branes with non-Killing horizons dual to the NESSs are constructed following the method outlined in \cite{Figueras:2012rb}, where the Einstein-DeTurck equations \cite{Headrick:2009pv, Wiseman:2011by}
\be
R_{ab} - \nabla_{(a}\xi_{b)} + 3 g_{ab} = 0
\ee
are solved with a coordinate system specified by a reference metric that penetrates the future event horizon.
The vector $\xi^{a} \equiv g^{bc}(\Gamma^a_{bc} - \bar{\Gamma}^a_{bc})$, where $\bar{\Gamma}$ is the connection for a reference metric, here chosen to be
\be
\bar{ds}^2 = ds_{Schw.}^2 + z^{-2} s_{\mu\nu}(x) dx^\mu dx^\nu. \label{refmetric}
\ee
Now $u^\mu$ is given by the 2 parameters $\beta^i$ (instead of $v^i$).
Our method differs in places from that given in \cite{Figueras:2012rb} because we include flow transverse to the obstacle and cases which are asymptotically supersonic. Because of these differences we go into some detail of the method in this section. 
 See also \cite{Fischetti:2012vt} for other non-Killing black hole constructions.
 
The obstacle is provided by gradients in the boundary metric.
We consider sources of the form, $s_{\mu\nu} = \Delta_\mu^{~\rho}\Delta_\nu^{~\sigma}\mathcal{S}_{\rho\sigma}$, so that $u^\mu s_{\mu\nu} = 0$.  For concreteness we adopt a particularly simple choice of source, $s_{\mu\nu} = \mathcal{S}_{\mu\nu} = s(x) n_\mu n_\nu$ where $n_\mu = \left(\beta_x^2+\beta_y^2\right)^{-1/2}(0, -\beta_y, \beta_x)$ and $s(x) = A e^{-B x^2}$.
For the physical metric we factor out a boundary-divergent term,
\be
ds^2 = g_{ab} dx^a dx^b = \frac{1}{z^2}h_{ab}(z,x) dx^a dx^b,
\ee
and since we also consider the case of shear flow we keep all 10 metric components.

The system is extended and inhomogeneous in the $x$ direction, and so we compactify it using instead a coordinate $\rho$, 
\be
x = \frac{\rho/\ell}{1-\rho^2}
\ee
where $\rho$ goes from $-1$ to $+1$ and where $\ell$ is a parameter which will allow the stretching of the coordinates relative to any characteristic feature size. We adopt a regular grid taking $N_z$ points in the $z$ coordinate and $N_\rho = 4 N_z$ points in the $\rho$ coordinate. The radial coordinate goes from $z=0$ at the boundary to $z=1$ which we demand is situated behind an event horizon. We utilise sixth-order finite differences for both the $z$ and $\rho$ derivative operators.

At $\rho=\pm 1$ we impose Neumann boundary conditions on all variables. At $z=0$ we impose Dirichlet boundary conditions on all variables, specifically we fix each component in terms of the reference metric, $(h_{ab})_{z=0} = (z^2 \bar{g}_{ab})_{z=0}$. This Dirichlet boundary condition fixes the boundary metric but it does not fix enough data to uniquely specify a solution, since there is a moduli space of flowing solutions where one can vary the energy density and velocity. Thus, following \cite{Figueras:2012rb} we fix further data using points at $z=1$, behind the event horizon. All other points evolve freely according to the equations of motion. For the solutions with transverse flow we have three moduli, and so we equate $g_{ab}$ to $\bar{g}_{ab}$ at three further points behind the event horizon at $z=1$. This prescription varies depending on whether the flow is asymptotically subsonic or supersonic. In detail, the pattern of moduli fixing is performed as follows,
\bgroup
\begin{center}
\begin{tabular}{@{\hspace{1em}}c@{\hspace{1em}}|@{\hspace{1em}}c@{\hspace{1em}}|@{\hspace{1em}}c@{\hspace{1em}}}
asymptotic $v$ &  $vx=-\infty$  &  $vx=+\infty$\\
\hline
subsonic & $g_{tt}, g_{ty}$ & $g_{tz}$\\
supersonic & $g_{tt}, g_{tz}, g_{ty}$ & --
\end{tabular},
\end{center}
\egroup
thereby ensuring that the parameters of the reference metric ($\beta_x, \beta_y, z_h$) determine the eventual moduli of the solution, $(\varepsilon, v, \theta)_{L,R}$.
Note that the supersonic case corresponds to a complete specification of these moduli on the upstream side. The result is a set of boundary conditions -- and correspondingly a set of solutions -- which are completely specified by the parameters in the reference metric, i.e. $\beta_x, \beta_y, z_h$, together with the source function.

The discretised equations with boundary conditions as described, are iteratively solved using the Newton method which is continued until a threshold residual is met everywhere on the grid. An initial guess metric for this iterative process is $\bar{g}$. Initially low resolution solutions are constructed, around $N_z=20$ and these are then used as an initial guess for a higher resolution solution, with sixth-order interpolation used to generate values at the new grid points. Typically only 1-2 Newton steps are required in this process to meet the residual threshold, making it very efficient for reaching higher resolutions. We check convergence of the $\xi^\mu$ vector to zero with increasing resolution. We use \verb|binary64| floating-point format for our numerics.

\begin{figure}
\vspace{2em}
\includegraphics[width=\columnwidth]{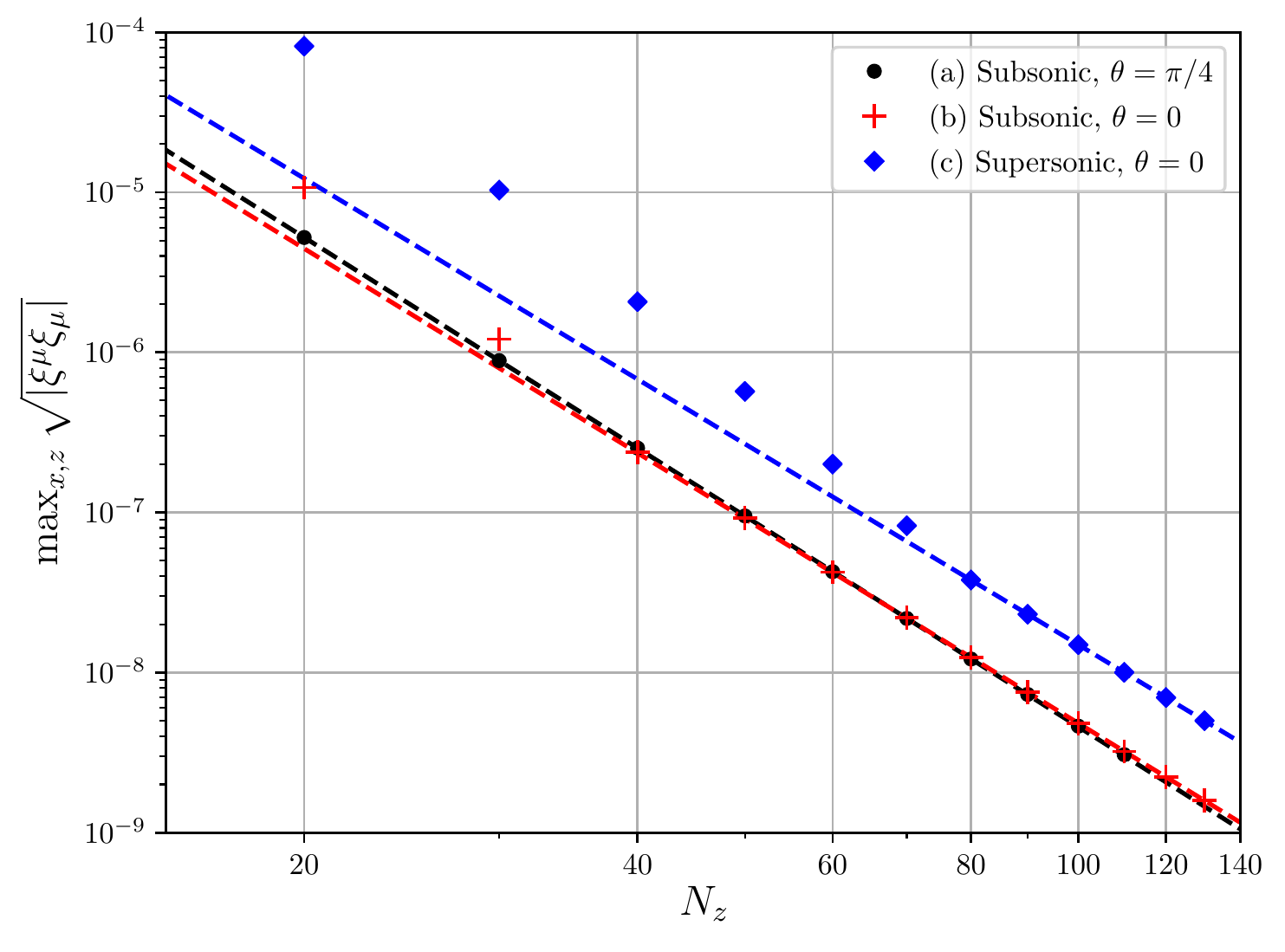}
\caption{Convergence of $\xi^\mu$, towards zero in the continuum limit, for a grid of size $N_z\times 4N_z$. We show the maximum absolute value of $|\xi|$ on the numerical grid, excluding $z=1$ points behind the horizon. Power-law convergence is given by straight lines on this log-log plot with best fits given by the dashed lines (using the six largest $N_z$ points). For the solutions as labelled we have approximately fourth-order convergence, with the best fit rates: (a) $4.4$, (b) $4.2$, (c) $4.2$.}
\label{Fig.DeTurck}
\end{figure}

The solutions used in the main text correspond to the following parameter sets,
\begin{enumerate}[label=(\alph{enumi})]
\item Fig. 3 main text, subsonic-to-subsonic with transverse flow: $z_h =  0.975, \beta_x = 0.15, \beta_y = 0.15, A= 1.0, B = 3.0, \ell=0.5$.
\item Fig. 4 main text, lower left panel, subsonic-to-subsonic without transverse flow: $z_h =  0.975, \beta_x = 0.6, \beta_y = 0.0, A= 0.1, B = 2.9618, \ell=0.5$. 
\item Fig. 4 main text, lower right panel, supersonic-to-supersonic without transverse flow: $z_h =  0.975, \beta_x = 0.8, \beta_y = 0.0, A= 0.1, B = 3.0, \ell=0.5$.
\end{enumerate}
For each of these solutions the convergence of $\xi^\mu$ towards zero is shown in Fig. \ref{Fig.DeTurck}. We find approximate fourth-order convergence, consistent with the overall convergence of the solutions found in \cite{Figueras:2012rb} which also used sixth order finite differences.

Near the AdS boundary at $z=0$ we may relate the coefficients of the expansion of the various metric functions to one-point functions of the stress tensor in the dual field theory. In particular we solve the Einstein equations in a near-boundary Fefferman-Graham gauge, where we may read off the dual one-point functions after holographic renormalisation \cite{deHaro:2000vlm}. We then convert to the coordinates used defined by $\xi^\mu=0$, used in this paper. The stress tensor can be extracted by taking three z-derivatives of various metric functions. For instance, 
\be
\left<T_{\mu\nu}\right> = \frac{1}{2}\partial_z^3 h_{\mu\nu}\big|_{z=0} + \frac{\gamma_{\mu\nu}}{z_h^3} + V_{\mu\nu}
\ee
where the $\left<T_{\mu\nu}\right>$ denotes the expectation value of the CFT stress tensor, and $V_{\mu\nu}$ is a set of terms, included in our analysis, but which we have omitted for this presentation. Furthermore these vanish when $\partial_x s_{\mu\nu} = \partial_x^2 s_{\mu\nu} = \partial_x^3 s_{\mu\nu} = 0$. Such terms could be omitted outside the source region when looking to extract the modes of interest. We note that it is sometimes not possible to solve the eigenvalue problem \eqref{evalproblem} everywhere along the flow, depending on how strongly the obstacle deforms the flow. However solutions to \eqref{evalproblem} always exist at large enough distances since the flow returns to equilibrium.

\end{document}